# Disposable Opto-Acoustic Window Enabled Cost-effective Photoacoustic-Ultrasound Dual-modal Imaging

Yunhui Jiang, [1] Fan Zhang, [1] Yuwei Zheng, [1] Ruixi Sun, [1] and Fei Gao[1,2,3,*]

1 Hybrid Imaging System Laboratory, School of Information Science and Technology, ShanghaiTech University, 393 Middle Huaxia Road, Shanghai, 201210

2 School of Biomedical Engineering, Division of Life Sciences and Medicine, University of Science and Technology of China, Hefei, Anhui, 230026, China

3 Suzhou Institute for Advanced Research, University of Science and Technology of China, Suzhou, Jiangsu, 215123, China

\* xjtugaofei@foxmail.com

*Abstract*

Photoacoustic imaging (PAI) and ultrasound imaging (USI) are important biomedical imaging techniques, due to their unique and complementary advantages in tissue's structure and function visualization. In this Letter, we proposed a coaxial photoacoustic-ultrasound dual-modal imaging system (coPAUS) with disposable opto-acoustic window. This opto-acoustic window allows part of light to go through it, and another part of light to be converted to ultrasound transmission signal by photoacoustic effect. By single laser pulse illumination, both PA signals and reflected US signals can be generated. Then, a linear array probe receives both PA and US signals, enabling simultaneous dual-modal PA and US imaging. Ex vivo experiments were conducted involving pencil lead, hair, and plastic tube with black spot, as well as in vivo experiment on human finger. The system's resolutions for PA and US imaging are 215 µm and 91.125 µm, with signal-to-noise ratios for PA and US signals reached up to 37.48 dB and 29.75 dB, respectively, proving the feasibility of the coPAUS dual-modal imaging. The proposed coPAUS system with disposable opto-acoustic window provides an immediate and cost-effective approach to enable US imaging capability based on an existing PA imaging system.

*Introduction*

PAI, a non-invasive biomedical imaging technique based on the photoacoustic effect, reveals endogenous chromophores (such as hemoglobin and melanin) and exogenous contrast agents (such as metal nanoparticles and organic molecules) by generating ultrasound signals through transient thermoelastic expansion induced by laser excitation. High-resolution images are reconstructed normally using delay-and-sum (DAS) algorithms [1-5]. Photoacoustic tomography (PAT) can achieve sub-millimeter resolution at depths of several centimeters, providing the optical contrast and functional imaging of tissues [6]. USI, on the other hand, acquires morphological information of tissues through the propagation of sound waves and their reflection at interfaces with different acoustic impedances. The two imaging modalities offer complementary strengths in terms of morphological and functional imaging [7].

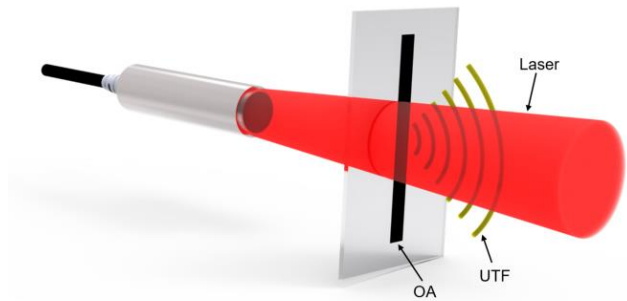

Fig. 1. Schematic diagram of OAW enabling co-axial laser illumination and ultrasound transmission field. OA, Optical absorber; UTF, Ultrasound transmission field.

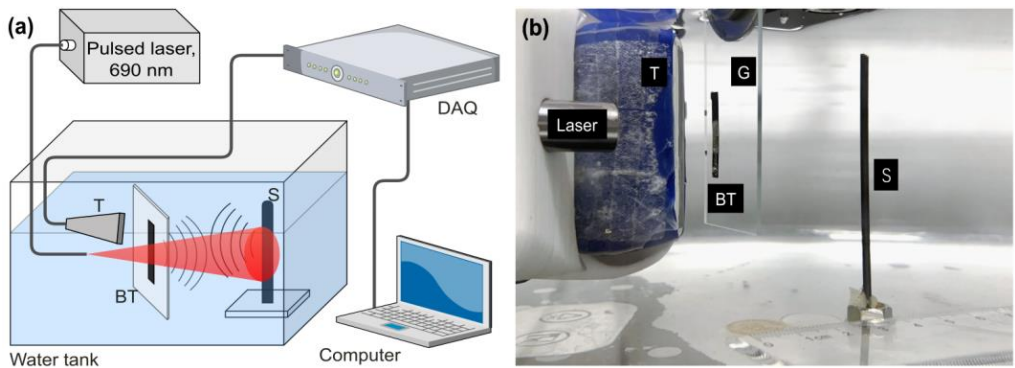

Fig. 2. (a) Experimental setup of coPAUS system. (b) Photograph of coPAUS system. T, Transducer; DAQ, Data Acquisition Card; BT, Black Tape; S, Sample; G, Glass.

Traditional PAUS dual-modal imaging systems often integrate an external ultrasound transmission module into a PA imaging system, which usually suffers higher cost, complex synchronization design, and un-coaxial light/sound illumination [8-12]. In this Letter, we propose co-axial photoacoustic-ultrasound (coPAUS) dual-modal imaging with disposable opto-acoustic window (OAW). A conceptual illustration of the OAW is shown in Fig. 1, which enables simultaneous light/sound illumination by a single pulsed laser. Specifically, part of the laser energy is transmitted through the window to illuminate the imaging target, the other part of the laser energy is absorbed by the optical absorber patterned on the window [13-16]. The unique advantage of OAW is its feasibility of generating custom-designed ultrasound transmission field at extremely low cost. Such simple and low-cost feature of OAW also enables its disposable usage for switching different OAW for different imaging scenarios.

*Experiment*

In this work, we designed the OAW module by sticking a black tape on a glass slide, which can be flexibly positioned in front of the laser output. The experimental setup is shown in Figure 2(a). It employs a pulsed laser (Lasersound-OPO-M-10, TsingPAI Technology, Pte Ltd) with a wavelength of 690 nm and a repetition rate of 10 Hz as the light source. The black tape is adhered to the glass slide and positioned in front of the laser beam path. To optimize laser energy utilization, the width of the black tape is designed to be smaller than the diameter of the laser beam. Due to significant optical absorption of the black tape, it generates a strong laser-induced ultrasound (LUS) signal based on PA effect. In the experiment, both the transmitted laser through the glass, and the LUS signal generated by the black tape, propagate through water. The round-trip LUS signal reflects off the object's surface, together with the single-trip PA signals generated by the object through PA effect [17]. These signals are ultimately received by a linear array probe with a central frequency of 7.5 MHz. The signals are acquired at a sampling rate of 40 MSPS using a high-speed data acquisition card (HISonics, HIS PATech Pte. Ltd.) with a total of 4096 sampling points, which are transmitted in real-time to a computer for image reconstruction. The photograph of the experimental setup is illustrated in Figure 2(b).

To validate the feasibility of coPAUS dual-modal imaging system, we used phantoms including a 2-mm-diameter pencil lead, a 0.07-mm-diameter human hair, and a 5.2-mm-diameter white plastic tube with black spot. Figure 3(a) displays the generated PA and reflected US signals from these phantoms. The waveforms reveal two distinct signal peaks in time domain, with the US signal's propagation time approximately twice that of the PA signal. This difference arises because, in PAI, the laser's propagation time in water is negligible, and only single-trip propagation of the received PA signal is considered. In contrast, US signal involves round-trip propagation time for both transmission and receiving.

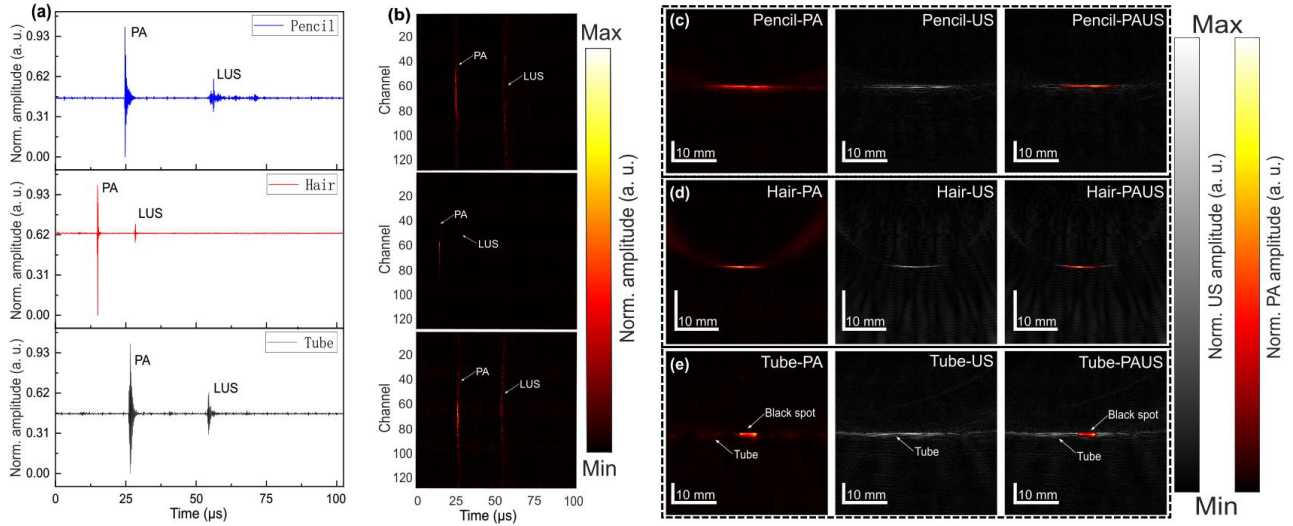

Fig. 3. (a) The received PA and US signals from pencil, hair, and a tube with black dot. (b) The sinogram images of the three phantoms mentioned above. (c)-(e) The PA, US, and PA-US imaging results of the three phantoms.

Figure 3(b) presents the sinogram images of PA and US signals, illustrating the distribution of signals captured by different channels of the linear probe over the sampling period. Due to the higher laser intensity at the central part, both PA and US signal's intensity is also concentrated at the center of the US probe and along the laser beam path. Additionally, the signal intensity of the US signal is obviously lower than the PA signal, which is caused by the longer propagation distance and attenuation of US signal. Figures 3(c)-(e) show the reconstructed images of PA, US, and their overlapped PAUS dual-modal images. For the pencil lead phantom, both PAI and USI clearly delineate the shape and edges of the pencil. The PA image shows that pencil lead's length is approximately 2.3 cm, while the US image shows its length is roughly 3.84 cm. This demonstrate the better capability of recovering morphological information of USI. In imaging the hair, the coPAUS system effectively resolves fine structures. When imaging the white plastic tube with black spot, the plastic tube's weak light absorption at a wavelength of 690 nm results in strong PA signals only at the black spot, whereas the US image clearly outlines the plastic tube's contour. This demonstrates the dual-contrast advantages of the system: PAI to reveal optical absorption contrast, USI to reveal acoustic impedance contrast.

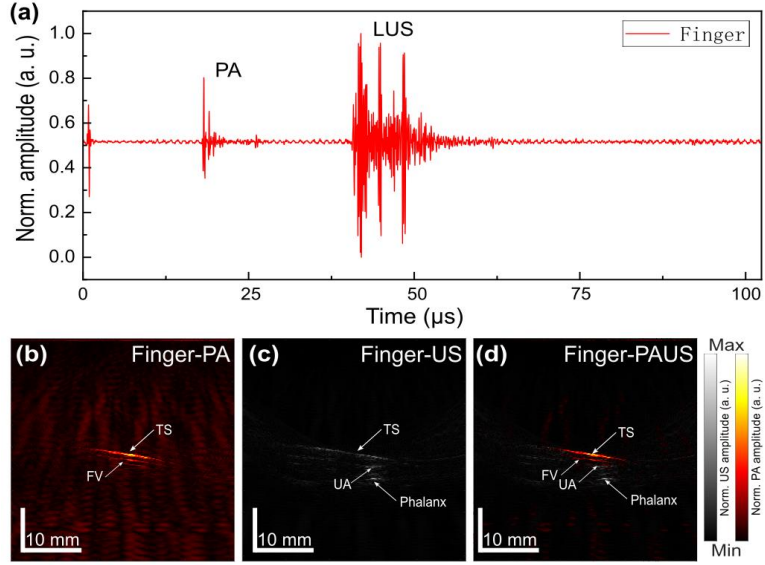

Fig. 4. (a) The received photoacoustic and ultrasound signals from the finger. (b)-(d) The PA, US, and PA-US imaging results of the finger. TS, tissue surface; FV, finger vasculature; UA, the upper aponeurosis.

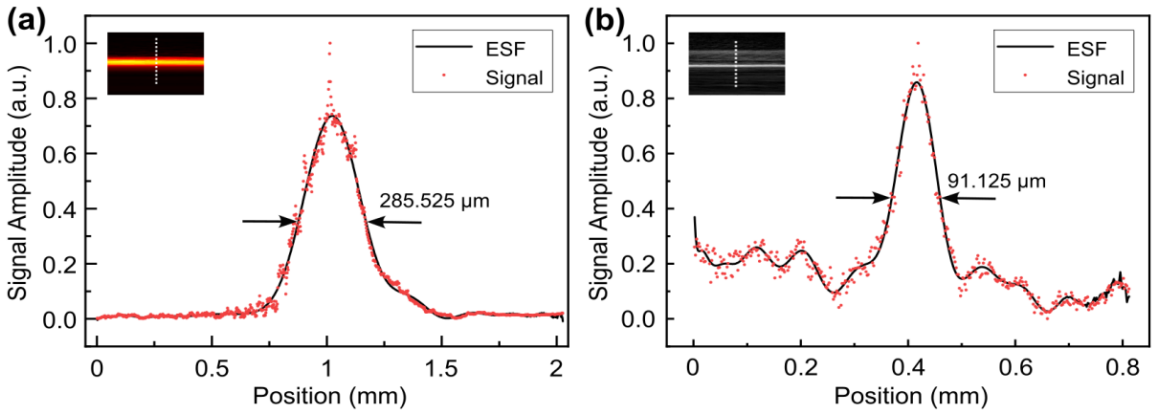

Fig. 5. Evaluation of the resolution of the coPAUS system: (a) resolution for photoacoustic imaging of a hair. (b) The resolution for ultrasound imaging of the pencil lead. ESF, edge spread function.

To further evaluate the feasibility of the proposed coPAUS system for in vivo biomedical imaging, we performed dual-modal imaging on the fingers of healthy volunteers, as depicted in Figure 4. The finger is an ideal subject for PAUS dual-modal imaging because it includes both vascular and bones, which are optimal endogenous contrast for PA and US imaging, respectively.

Additionally, given the high prevalence and significant clinical impact of finger arthritis, this also holds important clinical relevance [18]. In Figure 4(a), the PA signal was first detected by the linear probe at 17.75 μs, while the LUS signal was received after 22.71 μs.

Figures 4(b)-(d) illustrate the reconstructed images of PAI and USI. In Figure 4(b), the PA imaging clearly reveals the skin surface and vascular structures of the finger, indicating high optical absorption properties of the skin and blood vessels, though the PA contrast for the bone is relatively low. Figure 4(c) shows high acoustic contrast of the bone relative to surrounding tissues through US imaging, effectively outlining the surface contours of the deep subcutaneous finger bone. When PA and US images are fused, as shown in Figure 4(d), the relative positions of the finger's vascular and bony structures are distinctly visible.

Table 1. The signal-to-noise ratio of the PA/US signals for the four phantoms in the experiment.

| Unit: dB | Pencil lead | Hair | Tube | Finger |
| --- | --- | --- | --- | --- |
| PA | 31.73 | 37.48 | 32.27 | 23.48 |
| US | 20.15 | 23.44 | 22.89 | 29.75 |

In order to delineate the system's resolution, we analyzed the edge response for photoacoustic and ultrasound imaging of a hair strand and a pencil lead. As shown in Figure 5, the normalized experimental data were fitted using the edge spread function (ESF, black solid line). The full width at half maximum (FWHM) of the ESF, minus the diameter of the phantom, was used to represent the system's resolution. We observed that for imaging a 70 μm hair strand, the resolution of photoacoustic imaging was approximately 215 μm. For imaging a 2 mm pencil lead, the resolution of the ultrasound imaging was 91.125 μm. Although measurements at different positions along the phantom edge may slightly vary, and the performance of image reconstruction algorithm can influence resolution, this demonstrates that the coPAUS system is capable of identifying small objects with high resolution.

To characterize the quality of photoacoustic and ultrasonic signals produced by the system, we calculated the signal-to-noise ratio (SNR) of these signals, as shown in Table 1. The SNR of both US and PA signals exceed 20 dB in phantom and in vivo finger experiments, demonstrating the high signal quality of the proposed coPAUS system.

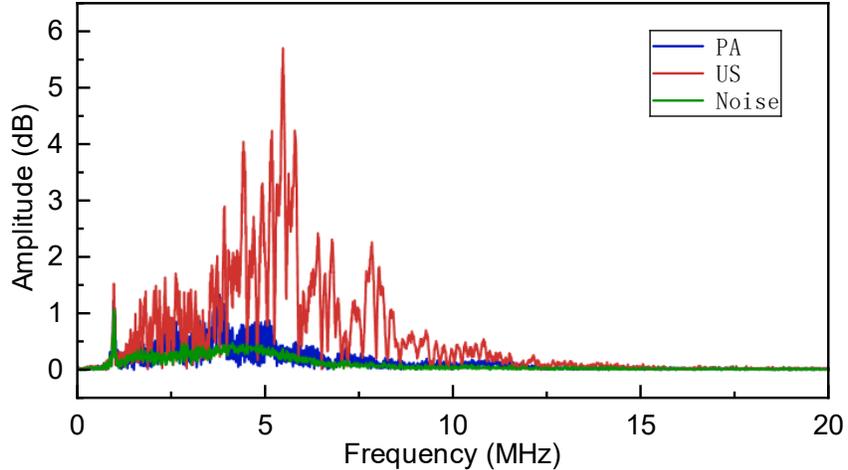

Fig. 6. The probe collected the spectrum of each component of the finger signal.

We also evaluated the signal's frequency response of the coPAUS system on in vivo finger. As shown in Figure 6, the power spectral density of both PA and US signals falls within 8 MHz. For PA signals, the primary peak of the spectrum is situated at 3.8 MHz with the maximum frequency reaching up to 7.5 MHz, limited by the bandwidth of the ultrasound probe. The spectrum range for the US signals is relatively broad, mainly due to the reflections and scatterings of the ultrasound signal on multiple tissue interfaces during transmission. The environmental and electronic noise from the system is primarily generated in the low-frequency range, with the maximum amplitude observed at 0.9 MHz, which can be well removed by proper filter design.

## *Conclusion*

In this letter, we propose a cost-effective coaxial photoacoustic-ultrasound dual-modal imaging system with disposable OAW. This system enables both PA and US imaging based on single-pulse excitation by just adding a very cheap OAW slide. In the future work, improvements can be achieved including the incorporation of fiber-optic beam-splitting to enable multi-angle illumination, replacement of linear transducers with bowl-shaped transducers to enable real-time 3D imaging, and custom-patterned OAW to optimize ultrasound transmission pattern for high-performance US imaging.